\documentclass[12pt,a4paper]{article}
\usepackage{a4wide}
\usepackage{amsmath}
\usepackage{amssymb}
\usepackage{epsfig}
\usepackage{subfigure}
\usepackage{exscale}
\usepackage{float}
\usepackage{bbm}
\usepackage[numbers,sort&compress]{natbib}
\usepackage{pst-plot, pstricks,pst-math}
\usepackage{fancybox,amssymb,color}
\usepackage{graphicx}
\usepackage{pstricks, color, graphicx, epsfig, psfrag}
\usepackage{amsfonts,amsmath,amssymb,slashed}
\usepackage{dsfont}
\usepackage{bbm,bm}
\usepackage{fancyhdr, a4wide}
\usepackage[english]{babel}
\usepackage{subfigure}
\makeatletter
\@addtoreset{equation}{section}
\makeatother

\title{Thermomagnetic Properties of QCD}

\author{Christoph P.\ Hofmann$^a$ \\ \\
\normalsize{$^a$ Facultad de Ciencias, Universidad de Colima} \\
\vspace{0.3cm}
\normalsize {Bernal D\'iaz del Castillo 340, Colima C.P.\ 28045, Mexico} \\}

\begin{document}

\maketitle

\begin{abstract} \normalsize

We explore the low-energy regime of quantum chromodynamics subjected to an external magnetic field by deriving the two-loop representations
for the entropy density and the magnetization within chiral perturbation theory (CHPT). At fixed temperature, the entropy density drops
when the magnetic field becomes stronger. The magnetization induced at finite temperature is negative in the entire parameter region
accessible by CHPT. We also point out that the enhancement of the finite-temperature part in the quark condensate is correlated with the
decrease of the entropy density.

\end{abstract}

\maketitle

\section{Introduction}
\label{Intro}

The thermodynamic properties of quantum chromodynamics in a homogeneous external magnetic field have been explored by many authors. In the
present study we focus on the entropy density and the magnetization -- and furthermore shed light on the connection between the quark
condensate and entropy density. Articles that also have discussed the dependence of entropy density and magnetization on temperature,
magnetic field strength, and pion mass are:
Refs.~\citep{BCLP10,BBES13,BBEKS14,BBES14,BCKMN19} based on lattice QCD,
Refs.~\citep{SMMR01,FS14,FTAPK17,AFPRT20} based on the Nambu-Jona-Lasinio model and extensions thereof, as well as
Refs.~\citep{Cha96,CW09,SWW10,End13,TPPYB15,SM16,PEP16,TDES16,RP17,TDH18,KGBHM19,LFL19,RP19,KB20} that rely on yet other methods.

Still, in the regime of low temperatures and weak magnetic fields, a comprehensive investigation of entropy density and magnetization
appears to be lacking. Here, within the framework of two-flavor chiral perturbation theory, we provide such a fully systematic analysis.
Based on earlier work of the author, Refs.~\citep{Hof19,Hof20a,Hof20b}, we derive the two-loop representations for the entropy density and
the magnetization.

We find that the entropy density, at fixed temperature, decreases when the magnetic field becomes stronger and it also decreases when the
masses of the pions grow. The impact of the magnetic field is most pronounced in the chiral limit. In the real world with pion masses fixed
at $M_{\pi} = 140 \, \text{MeV}$, the entropy density also drops in the presence of an external magnetic field whose impact is most distinct
around the temperature $T \approx 30 \, \text{MeV}$. While the results for the dependence of entropy density on magnetic field strength,
temperature, and {\it arbitrary} pion mass are new to the best of our knowledge, the magnetic-field induced decrease of the entropy density
at the physical point $M_{\pi} = 140 \, \text{MeV}$ has also been observed in the hadron resonance gas model \citep{End13}, and in the (2+1)
flavor Polyakov-loop quark-meson model \citep{LFL19}. However, the comparison is only qualitative because the latter reference is based on
three flavors, and the hadron resonance gas model includes even more particles. But most importantly, our CHPT study is fully systematic
and model independent.

The magnetization induced at finite temperature is negative in the entire parameter region accessible by CHPT
($T, M_{\pi} , \sqrt{|qH|} \lessapprox 0.2 \, \text{GeV}$), which includes the physically most relevant case $M_{\pi} = 140 \, \text{MeV}$.
The magnitude of the finite-temperature magnetization grows as both magnetic field strength and temperature increase, implying that the QCD
vacuum behaves as a {\it diamagnetic} medium at low temperatures and weak magnetic fields. This is fully consistent with conclusions drawn
from lattice QCD \citep{BEMNS14,BBEKS14,LT14,BEP20}, the (2+1) flavor Polyakov-loop quark-meson model \citep{LFL19}, and the three-flavor
quark-meson model with $U_A(1)$ anomaly \citep{KK15}. Still, the region accessible by CHPT has not been fully addressed in these
references.

Finally we point out that the characteristics of the entropy density and the finite-temperature quark condensate in a magnetic field are
correlated. This becomes most transparent when the $H$=0 portions in either quantity are subtracted to unmask the effect of the magnetic
field. We observe that for arbitrary pion masses -- including the physical point $M_{\pi} = 140 \, \text{MeV}$ -- the enhancement of the
finite-temperature quark condensate in a magnetic field is reflected in a decrease of the entropy density.

The article is organized as follows. To set the stage for our discussion, in Sec.~\ref{preliminaries} we provide the two-loop
representation for the free energy density and explain our notation. In Sec.~\ref{entropy} we derive the entropy density and explore its
dependence on temperature, magnetic field strength and arbitrary pion masses, including the physical point $M_{\pi} = 140 \, \text{MeV}$.
The two-loop representation for the magnetization induced at finite temperature is derived in Sec.~\ref{magnetization} and its properties
in magnetic fields and for arbitrary pion masses is elucidated in various figures. Sec.~\ref{quarkCondensate} is devoted to the connection
between order parameter and entropy density. Finally, Sec.~\ref{conclusions} contains our conclusions.

\section{Preliminaries}
\label{preliminaries}

Two-flavor chiral perturbation theory\footnote{Introductions to chiral perturbation theory are given in
Refs.~\citep{Leu95,Man97,Sch03,Goi04,Bra10}.} subjected to a magnetic background $H$ and at finite temperature has been used by various
authors to explore the low-energy regime of quantum chromodynamics
\citep{SS97,AS00,Aga00,AS01,CMW07,Wer08,AF08,And12a,And12b,BK17,Hof19,Hof20a,Hof20b}.
The starting point of the present study is the two-loop representation for the free energy density derived in
Ref.~\citep{Hof20a},\footnote{We confine ourselves to the isospin limit $m_u = m_d$.}
\begin{equation}
z = z_0 + z^T \, .
\end{equation}
Here $z_0$ is the vacuum energy density (free energy density at $T$=0) and $z^T$ represents the finite-temperature portion. The latter
amounts to
\begin{eqnarray}
\label{fedPhysicalM}
z^T & = & - g_0(M^{\pm}_{\pi},T,0) -\mbox{$ \frac{1}{2}$} g_0(M^0_{\pi},T,0)- {\tilde g}_0(M^{\pm}_{\pi},T,H) \nonumber \\
& & + \frac{M^2_{\pi}}{2 F^2} \, g_1(M^{\pm}_{\pi},T,0) \, g_1(M^0_{\pi},T,0)
- \frac{M^2_{\pi}}{8 F^2} \, {\Big\{ g_1(M^0_{\pi},T,0)  \Big\}}^2 \nonumber \\
& & + \frac{M^2_{\pi}}{2 F^2} \, g_1(M^0_{\pi},T,0) \, {\tilde g}_1(M^{\pm}_{\pi},T,H) + {\cal O}(p^8) \, ,
\end{eqnarray}
and relies on the kinematical Bose functions 
\begin{eqnarray}
g_0({\cal M},T,0) & = & T^4 \, {\int}_{\!\!\! 0}^{\infty}  \mbox{d} \rho \rho^{-3} \, \exp\Big( -\frac{{\cal M}^2}{4 \pi T^2}
\rho \Big) \Bigg[ S\Big( \frac{1}{\rho} \Big) -1 \Bigg] \, , \nonumber \\
g_1({\cal M},T,0) & = & \frac{T^2}{{4 \pi}} \, {\int}_{\!\!\! 0}^{\infty}  \mbox{d} \rho \rho^{-2} \, \exp\Big( -\frac{{\cal M}^2}{4 \pi T^2}
\rho \Big) \Bigg[ S\Big( \frac{1}{\rho} \Big) -1 \Bigg] \, , \nonumber \\
{\tilde g}_0(M^{\pm}_{\pi},T,H) & = & \frac{T^2}{{4 \pi}} \, |qH| {\int}_{\!\!\! 0}^{\infty} \mbox{d} \rho \rho^{-2} \,
\Bigg( \frac{1}{\sinh(|qH| \rho /4 \pi T^2)} - \frac{4 \pi T^2}{|qH| \rho} \Bigg) \nonumber \\
& & \times \, \exp\Big( -\frac{{(M^{\pm}_{\pi})}^2}{4 \pi T^2} \rho \Big) \Bigg[ S\Big( \frac{1}{\rho} \Big) -1 \Bigg] \, , \nonumber \\
{\tilde g}_1(M^{\pm}_{\pi},T,H) & = & \frac{1}{16 \pi^2} \, |qH| {\int}_{\!\!\! 0}^{\infty} \mbox{d} \rho \rho^{-1} \,
\Bigg( \frac{1}{\sinh(|qH| \rho /4 \pi T^2)} - \frac{4 \pi T^2}{|qH| \rho} \Bigg) \nonumber \\
& & \times \, \exp\Big( -\frac{{(M^{\pm}_{\pi})}^2}{4 \pi T^2} \rho \Big) \Bigg[ S\Big( \frac{1}{\rho} \Big) -1 \Bigg] \, ,
\end{eqnarray}
with $S(z)$,
\begin{equation}
S(z) = \sum_{n=-\infty}^{\infty} \exp(- \pi n^2 z) \, ,
\end{equation}
as the Jacobi theta function. These kinematical Bose functions depend on the masses of the charged ($M^{\pm}_{\pi}$) and neutral ($M^0_{\pi}$)
pions in a magnetic field, namely,
\begin{eqnarray}
\label{chargedNeutralPionMass}
{(M^{\pm}_{\pi})}^2 & = & M^2_{\pi} + \frac{{\overline l}_6 - {\overline l}_5}{48 \pi^2} \, \frac{{|qH|}^2}{F^2} \, , \nonumber \\
{(M^0_{\pi})}^2 & = & M^2_{\pi}  + \frac{M^2}{F^2} \, K_1 \, ,
\end{eqnarray}
where $K_1$ corresponds to the integral
\begin{equation}
\label{intK1}
K_1 = \frac{|qH|}{16 \pi^2}  \, {\int}_{\!\!\! 0}^{\infty} \mbox{d} \rho \, \rho^{-1} \, \exp\Big( -\frac{M^2_{\pi}}{|qH|} \rho \Big) \,
\Big( \frac{1}{\sinh(\rho)} - \frac{1}{\rho} \Big) \, ,
\end{equation}
$q$ is the electric charge, and ${\overline l}_5, {\overline l}_6$ are renormalized next-to-leading order (NLO) low-energy effective
constants. The mass $\cal M$ appearing in the kinematical Bose functions $g_0$ and $g_1$ can either stand for $M^{\pm}_{\pi}$ or $M^0_{\pi}$.
Finally, the mass $M_{\pi}$ is the renormalized pion mass in {\it zero} magnetic field,
\begin{equation}
\label{Mpi}
M^2_{\pi} = M^2 - \frac{{\overline l}_3}{32 \pi^2} \, \frac{M^4}{F^2} + {\cal O}(M^6) \, ,
\end{equation}
where $M$ ($F$) is the tree-level pion mass (pion decay constant).

\section{Entropy Density}
\label{entropy}

In the previous section we have defined the kinematical functions $g_r$ and ${\tilde g}_r$ that are dimensionful. In what follows it is
more convenient to use the dimensionless functions $h_r$ and ${\tilde h}_r$,
\begin{equation}
\label{conversion}
h_0 = \frac{g_0}{T^4} \, , \quad  {\tilde h}_0 = \frac{{\tilde g}_0}{T^4} \, , \qquad
h_1 = \frac{g_1}{T^2} \, , \quad  {\tilde h}_1 = \frac{{\tilde g}_1}{T^2} \, , \qquad
h_2 = g_2 \, , \quad  {\tilde h}_2 = {\tilde g}_2 \, .
\end{equation}
In addition, instead of using absolute values of temperature, pion mass and magnetic field strength, we prefer to work with the normalized
and dimensionless quantities $t, m$, and $m_H$ defined as
\begin{equation}
t = \frac{T}{4 \pi F} \, , \qquad m = \frac{M_{\pi}}{4 \pi F} \, , \qquad m_H = \frac{\sqrt{|q H|}}{4 \pi F} \, .
\end{equation}
The common denominator represents the chiral symmetry breaking scale $\Lambda_{\chi} \approx 4 \pi F \approx 1 \text{GeV}$. In the
low-energy region where chiral perturbation theory operates, the parameters $t, m$, and $m_H$ are small. In subsequent plots, the value of
the (tree-level) pion decay constant is $F = 85.6 \, \text{MeV}$ (see Ref.~\citep{Aok20}).

The entropy density $s$ can be extracted from the pressure via
\begin{equation}
s = \frac{\mbox{d} P}{\mbox{d} T} \, .
\end{equation}
The pressure in a homogeneous medium, up to the sign, is nothing but the finite-temperature piece in the free energy density,
\begin{equation}
P = -z^T \, .
\end{equation}
Derivatives of the Bose functions $g_r$ and ${\tilde g}_r$ with respect to temperature are easily obtained with the relations
\begin{eqnarray}
\frac{\mbox{d}}{\mbox{d} T} \, g_r({\cal M},T,0) & = & \frac{2 {\cal M}^2}{T} \, g_{r+1}({\cal M},T,0) + \frac{d-2r}{T} \, g_r({\cal M},T,0)
\, , \nonumber \\
\frac{\mbox{d}}{\mbox{d} T} \, {\tilde g}_r(M^{\pm}_{\pi},T,H) & = & \frac{2 M^2}{T} \, {\tilde g}_{r+1}(M^{\pm}_{\pi},T,H)
+ \frac{d-2r-2}{T} \, {\tilde g}_r(M^{\pm}_{\pi},T,H) \nonumber \\
& & + {\tilde g}^{[1]}_r(M^{\pm}_{\pi},T,H) \, , 
\end{eqnarray}
where
\begin{eqnarray}
{\tilde g}^{[1]}_r(M^{\pm}_{\pi},T,H) & = & \frac{T^{d-2r-2}}{{(4 \pi)}^{r+1}} \, |qH| {\int}_{\!\!\! 0}^{\infty} \mbox{d} \rho \rho^{r-\frac{d}{2}} \,
\Bigg( \frac{|qH| \rho \coth(|qH| \rho /4 \pi T^2)}{2 \pi T^3 \sinh(|qH| \rho /4 \pi T^2)} - \frac{8 \pi T}{|qH| \rho} \Bigg) \nonumber \\
& & \times \, \exp\Big( -\frac{{(M^{\pm}_{\pi})}^2}{4 \pi T^2} \rho \Big) \Bigg[ S\Big( \frac{1}{\rho} \Big) -1 \Bigg] \, .
\end{eqnarray}

On the basis of the representation for $z^T$, Eq.~(\ref{fedPhysicalM}), the two-loop entropy density -- scaled by $1/T^3$ -- amounts to
\begin{eqnarray}
\label{entropyDensity}
\frac{s}{T^3} & = & 2 s_{\pm} h_1(M^{\pm}_{\pi},T,0) + 4 h_0(M^{\pm}_{\pi},T,0) + s_0 h_1(M^0_{\pi},T,0) + 2 h_0(M^0_{\pi},T,0) \nonumber \\
& & + 2 s_{\pm} {\tilde h}_1(M^{\pm}_{\pi},T,H)+ 2 {\tilde h}_0(M^{\pm}_{\pi},T,H) + {\tilde h}^{[1]}_0(M^{\pm}_{\pi},T,H) \nonumber \\
& & - 8 \pi^2 m^2 \Big\{ 2 s_{\pm} h_2(M^{\pm}_{\pi},T,0) h_1(M^0_{\pi},T,0) + 4 h_1(M^{\pm}_{\pi},T,0) h_1(M^0_{\pi},T,0)
\nonumber \\
& & + 2  s_0 h_1(M^{\pm}_{\pi},T,0) h_2(M^0_{\pi},T,0) + 2 s_0 h_2(M^0_{\pi},T,0) {\tilde h}_1(M^{\pm}_{\pi},T,H) \nonumber \\
& & + 2 h_1(M^0_{\pi},T,0) {\tilde h}_1(M^{\pm}_{\pi},T,H) + 2 s_{\pm} h_1(M^0_{\pi},T,0) {\tilde h}_2(M^{\pm}_{\pi},T,H) \nonumber \\
& & + h_1(M^0_{\pi},T,0) {\tilde h}^{[1]}_1(M^{\pm}_{\pi},T,H) - s_0 h_1(M^0_{\pi},T,0) h_2(M^0_{\pi},T,0) \nonumber \\
& & - h_1(M^0_{\pi},T,0) h_1(M^0_{\pi},T,0) \Big\} \, ,
\end{eqnarray}
with coefficients
\begin{eqnarray}
s_{\pm} & = & \frac{m^2}{t^2} + \frac{({\overline l}_6 - {\overline l}_5) m^4_H}{3 t^2} \, , \nonumber \\
s_0 & = & \frac{m^2}{t^2} + \frac{m^2 m_H^2}{t^2} \, {\int}_{\!\!\! 0}^{\infty} \mbox{d} \rho \, \rho^{-1} \, \exp\Big( -\frac{m^2}{m^2_H} \rho \Big) \,
\Big( \frac{1}{\sinh(\rho)} - \frac{1}{\rho} \Big) \, .
\end{eqnarray}
The dimensionless functions ${\tilde h}^{[1]}_0(M^{\pm}_{\pi},T,H)$ and ${\tilde h}^{[1]}_1(M^{\pm}_{\pi},T,H)$ are
\begin{equation}
{\tilde h}^{[1]}_0(M^{\pm}_{\pi},T,H) = \frac{{\tilde g}^{[1]}_0(M^{\pm}_{\pi},T,H)}{T^3} \, , \qquad
{\tilde h}^{[1]}_1(M^{\pm}_{\pi},T,H) = \frac{{\tilde g}^{[1]}_1(M^{\pm}_{\pi},T,H)}{T} \, .
\end{equation}

\begin{figure}
\begin{center}
\hbox{
\includegraphics[width=8.0cm]{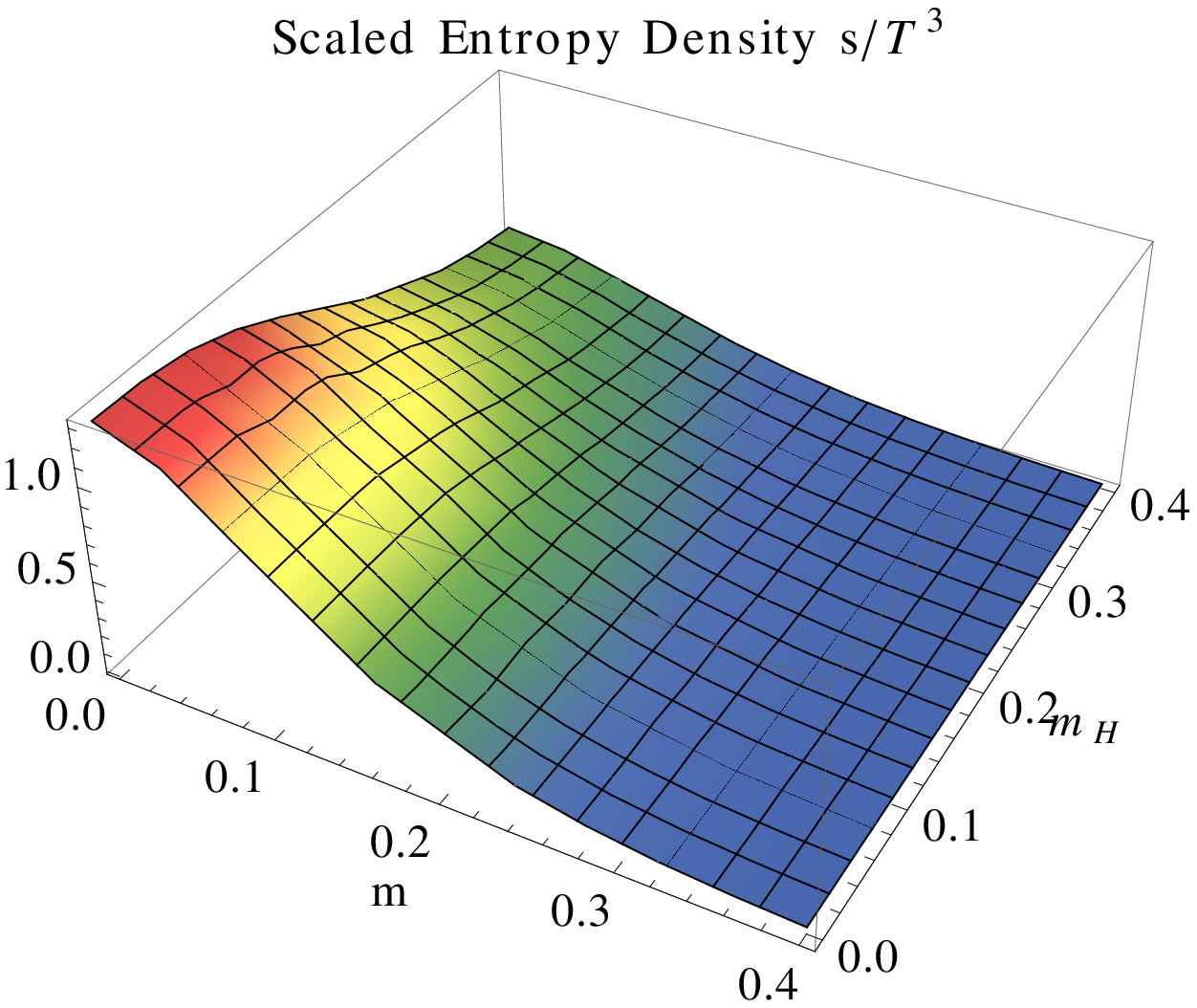}
\includegraphics[width=8.0cm]{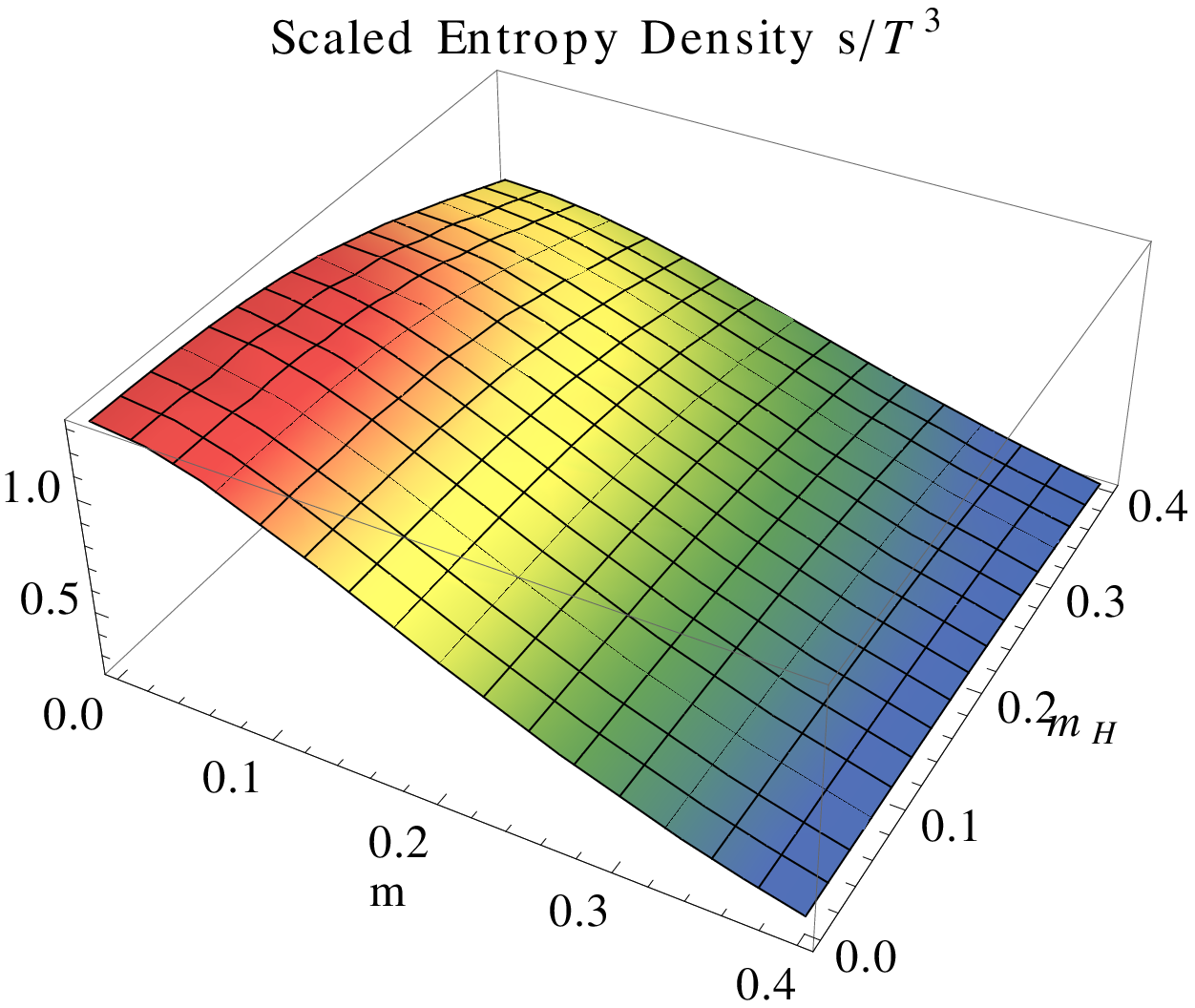}}
\end{center}
\caption{[Color online] Scaled entropy density $s/T^3$: Sum of one- and two-loop contributions for $T= 54 \, \text{MeV}$ (left) and
$T= 108 \, \text{MeV}$ (right) in terms of the dimensionless parameters $m,m_H$.}
\label{figure1}
\end{figure}

In Fig.~\ref{figure1} we plot the scaled entropy density $s/T^3$ -- sum of one- and two-loop contributions given in
Eq.~(\ref{entropyDensity}) -- in terms of magnetic field strength ($m_H$) and pion mass ($m$) at the fixed temperatures
$T= \{ 54, 108 \} \, \text{MeV}$ ($t= \{ 0.05, 0.1 \}$). The entropy density decreases when the magnetic field becomes stronger and it also
decreases when the masses of the pions grow. The impact of the pion mass, however, is more pronounced than the dependence on the magnetic
field. The limits $M \to 0$ (chiral limit) and $H \to 0$ do not pose any problems. In particular, taking the double limit $\{M, H\} \to 0$,
the scaled entropy density tends to the value describing the noninteracting Bose gas,
\begin{equation}
\frac{s(T,0,0)}{T^3} = \frac{2 \pi^2}{15} \approx 1.32 \, .
\end{equation}

\begin{figure}
\begin{center}
\hbox{
\includegraphics[width=8.0cm]{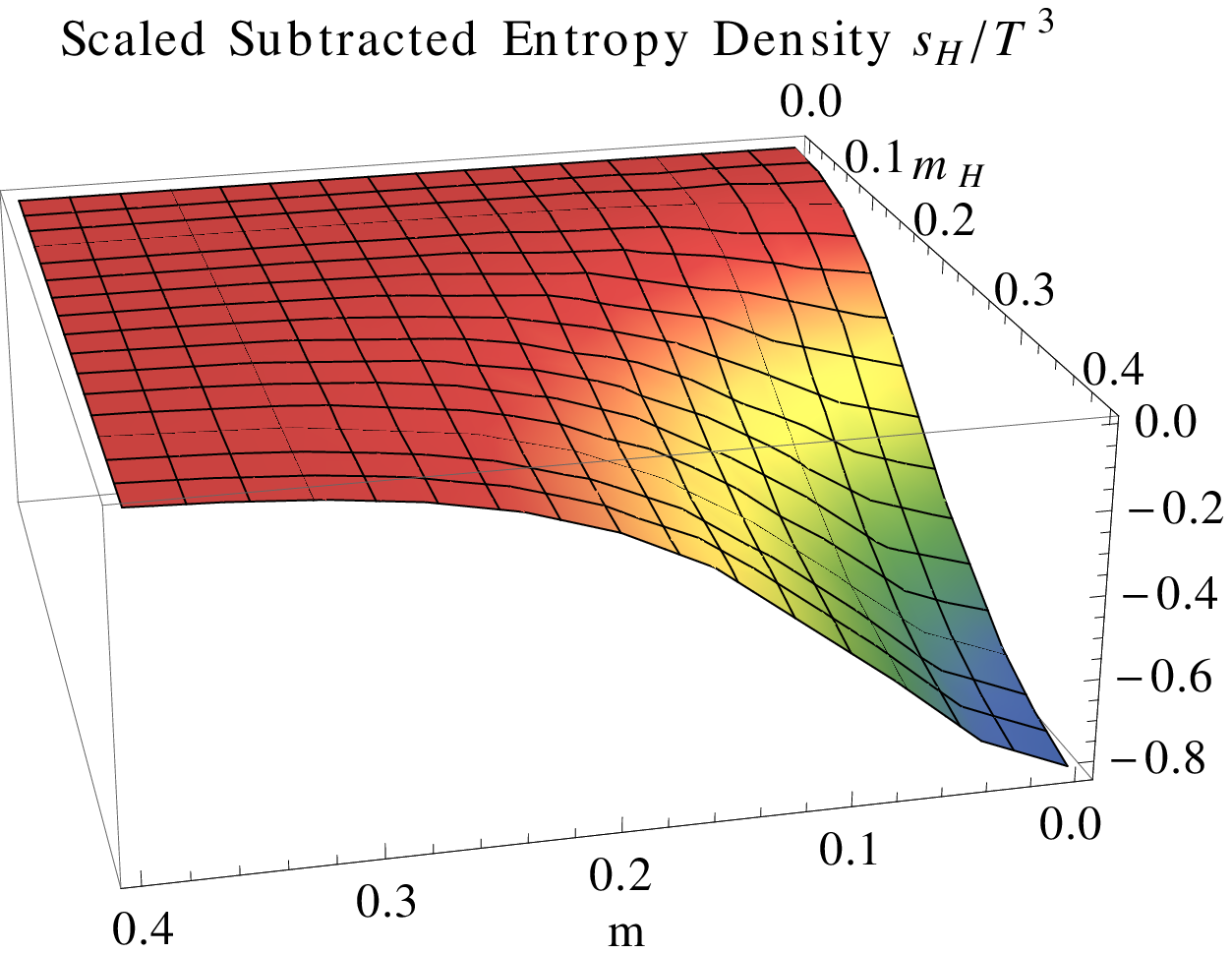}
\includegraphics[width=8.0cm]{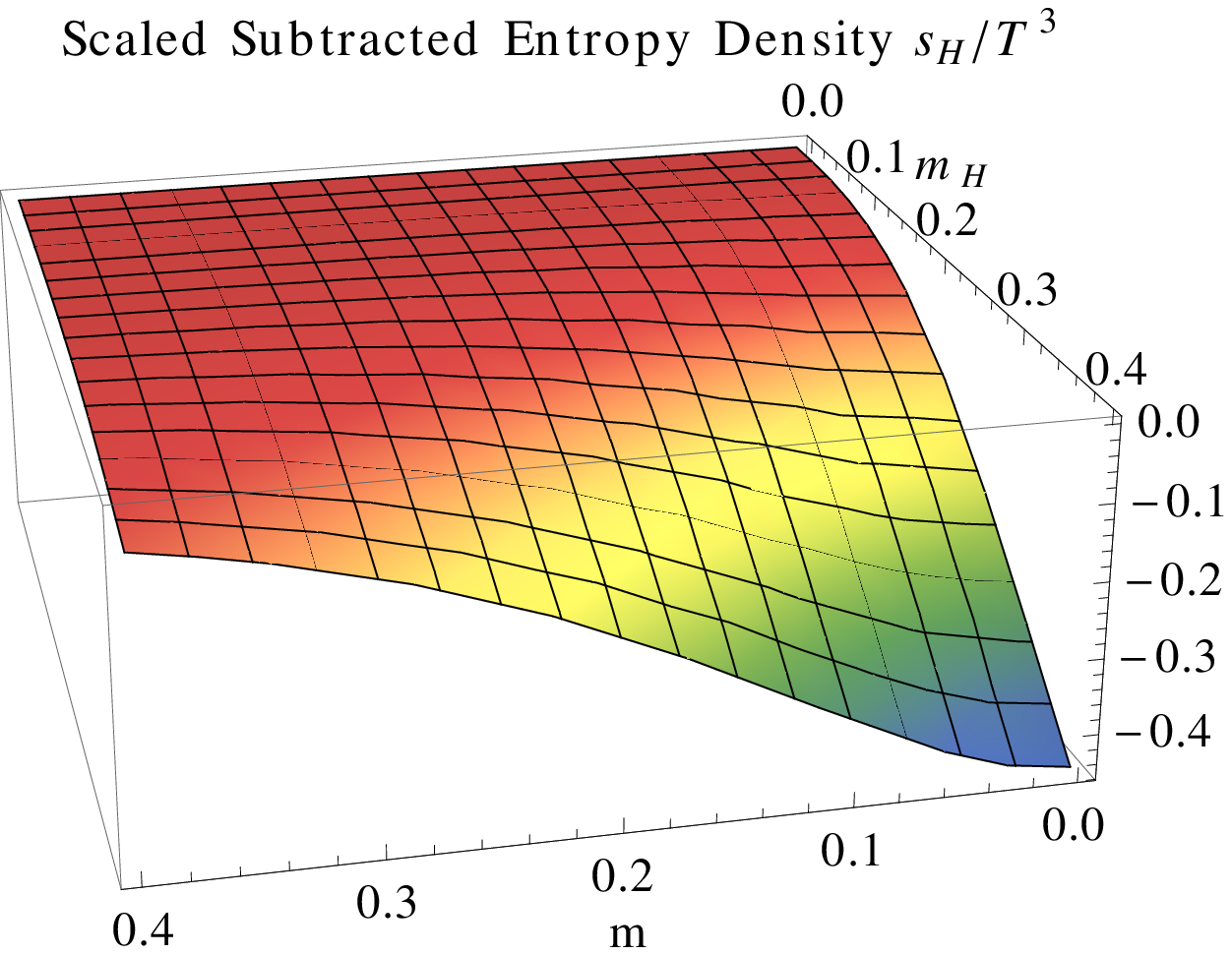}}
\end{center}
\caption{[Color online] Subtracted scaled entropy density  $s_H/T^3$: Sum of one- and two-loop contributions for $T= 54 \, \text{MeV}$
(left) and $T= 108 \, \text{MeV}$ (right) in terms of the dimensionless parameters $m,m_H$.}
\label{figure2}
\end{figure}

To better assess the effect of the magnetic field on entropy, we subtract the $H$=0 contribution, i.e., consider the quantity
\begin{equation}
\label{entropyH0subtracted}
\frac{s_H}{T^3} = \frac{s(T,M_{\pi},H) - s(T,M_{\pi},0)}{T^3}
\end{equation}
that measures the influence of the magnetic field. In Fig.~\ref{figure2} we plot $s_H/T^3$ in terms of magnetic field strength ($m_H$) and
pion mass ($m$) at the same fixed temperatures $T= \{ 54, 108 \} \, \text{MeV}$. Overall, in presence of the magnetic field, the entropy
density drops. For a given fixed $m_H$, the effect becomes most pronounced when the chiral limit is approached.

\begin{figure}
\begin{center}
\hbox{
\includegraphics[width=8.0cm]{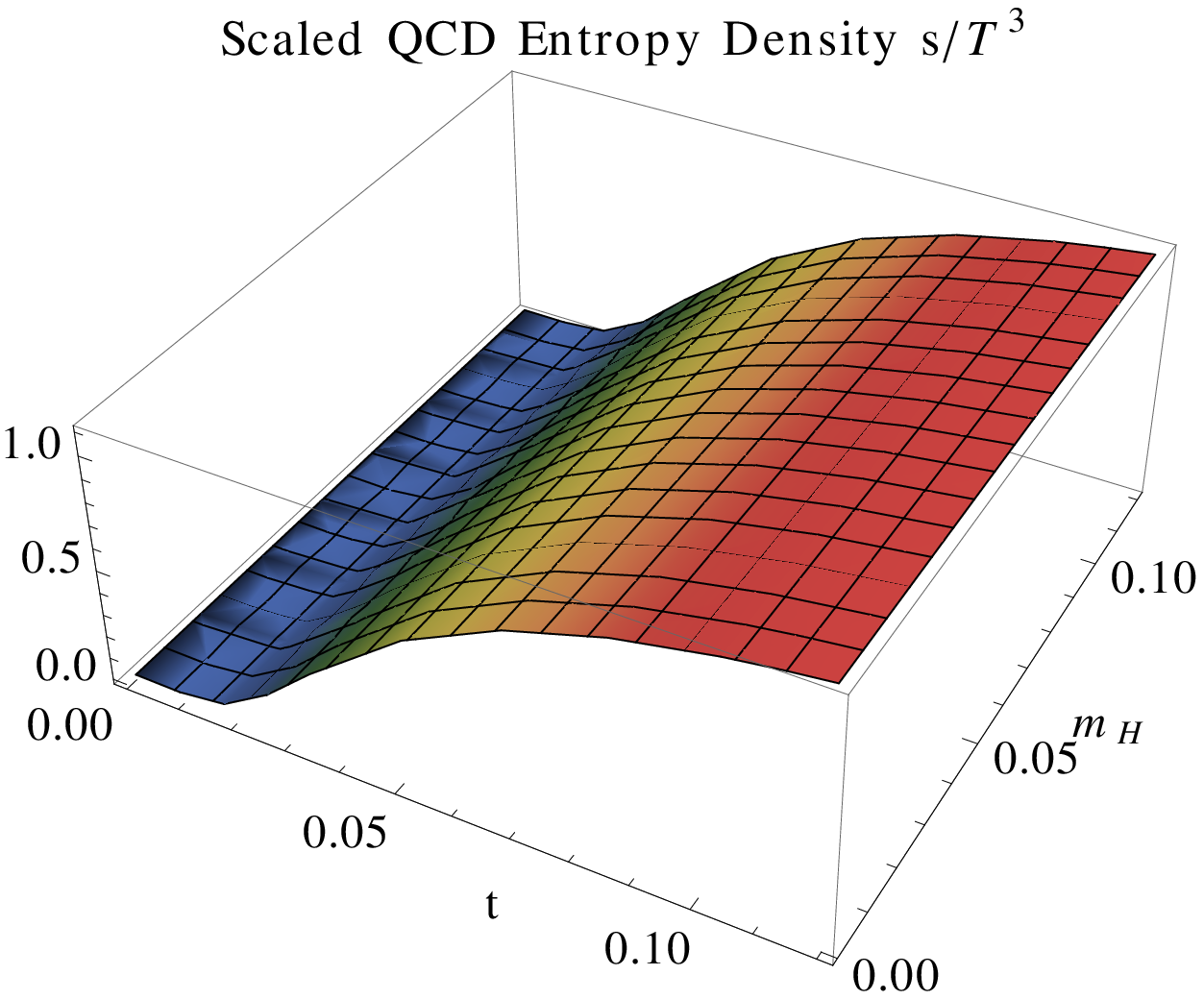}
\includegraphics[width=8.0cm]{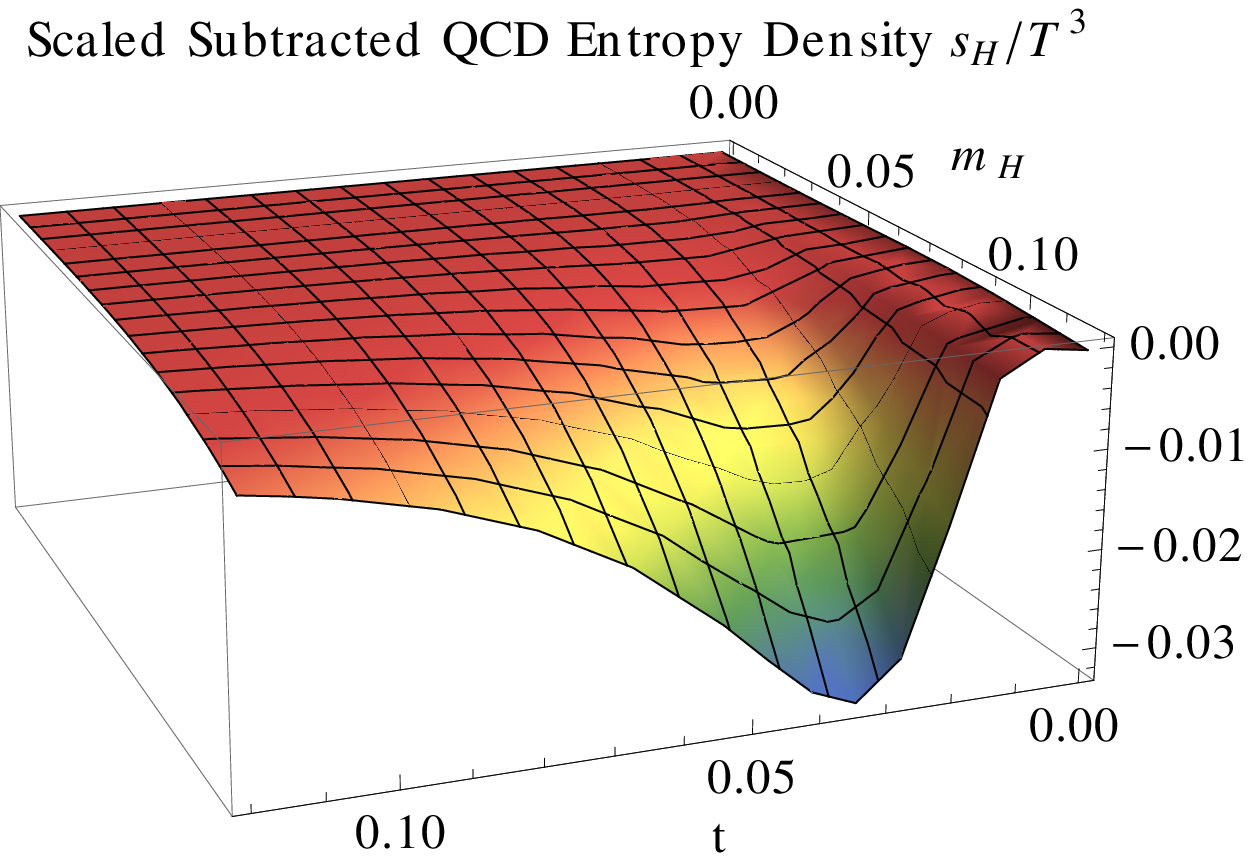}}
\end{center}
\caption{[Color online] Scaled QCD entropy density $s/T^3$ (LHS) and scaled subtracted QCD entropy density $s_H/T^3$ (RHS). Both quantities
in terms of the dimensionless parameters $m_H$ (magnetic field strength) and $t$ (temperature).}
\label{figure3}
\end{figure}

Let us now focus on the real world defined by the physical point $M_{\pi} = 140 \, \text{MeV}$ ($m = 0.130$). In Fig.~\ref{figure3}, on the
LHS, we plot the scaled two-loop entropy density $s/T^3$ according to Eq.~(\ref{entropyDensity}) in function of magnetic field strength
($m_H$) and temperature ($t$). Clearly, the impact of temperature is predominant and the slight decrease of the entropy density caused by
the magnetic field is hardly visible. Therefore, we rather consider the quantity $s_H/T^3$ defined by Eq.~(\ref{entropyH0subtracted}) that
isolates the effect of the magnetic field. Indeed, as illustrated on the RHS of Fig.~\ref{figure3}, the presence of the magnetic field
lowers the entropy density in an interesting and nontrivial way: the effect is most pronounced around $t \approx 0.035$, corresponding to
$T \approx 40 \, \text{MeV}$. As we discuss below, the increase of order -- as witnessed by the drop in entropy density -- is reflected in
the behavior of the finite-temperature quark condensate.

While our results regarding the dependence of entropy density on magnetic field strength, temperature, and {\it arbitrary} pion mass are
new to the best of our knowledge, the behavior of the entropy density at the physical pion mass has been explored in the hadron resonance
gas model \citep{End13}, as well as in the (2+1) flavor Polyakov-loop quark-meson model \citep{LFL19}. Indeed, both references also report
a decrease of entropy density caused by the magnetic field in the relevant region $T \lessapprox 100 \, \text{MeV}$ that is accessible with
chiral perturbation theory. The comparison, however, is only qualitative as the latter reference is based on three flavors and the hadron
resonance gas model also includes more particles than just the three pions.

\section{Magnetization}
\label{magnetization}

The magnetization ${\mathfrak M}$ is defined as the (negative) derivative of the free energy density with respect to the magnetic field,
\begin{equation}
{\mathfrak M}(T,M,H) = - \frac{\mbox{d} z}{\mbox{d} |qH|} \, .
\end{equation}
It contains both temperature-dependent and temperature-independent contributions. In this study we focus on the former.

The finite-temperature free energy density $z^T$, Eq.~(\ref{fedPhysicalM}), involves various types of kinematical Bose functions.
Corresponding derivatives with respect to the magnetic field are readily obtained via
\begin{eqnarray}
\frac{\mbox{d}}{\mbox{d} |qH|} \, g_r(M^{\pm},T,0) & = & -\frac{{\overline l}_6 - {\overline l}_5}{24 \pi^2} \, \frac{|qH|}{F^2} \,
g_{r+1}(M^{\pm},T,0) \, , \nonumber \\
\frac{\mbox{d}}{\mbox{d} |qH|} \, g_r(M^0,T,0) & = & -\frac{M^2}{F^2} \, \frac{\mbox{d} K_1}{\mbox{d} |qH|} \,
g_{r+1}(M^0,T,0) \, , \nonumber \\
\frac{\mbox{d}}{\mbox{d} |qH|} \, {\tilde g}_r(M^{\pm}_{\pi},T,H) & = & \frac{1}{|qH|} \, {\tilde g}_r(M^{\pm}_{\pi},T,H)
- \frac{{\overline l}_6 -{\overline l}_5}{24 \pi^2} \, \frac{|qH|}{F^2} \, {\tilde g}_{r+1}(M^{\pm}_{\pi},T,H) \nonumber \\
& & + {\tilde g}^{[H]}_r(M^{\pm}_{\pi},T,H) \, , 
\end{eqnarray}
where
\begin{eqnarray}
{\tilde g}^{[H]}_r(M^{\pm}_{\pi},T,H) & = & \frac{T^{d-2r-2}}{{(4 \pi)}^{r+1}} \, |qH| {\int}_{\!\!\! 0}^{\infty} \mbox{d} \rho \rho^{r-\frac{d}{2}} \,
\Bigg( -\frac{\rho \coth(|qH| \rho /4 \pi T^2)}{4 \pi T^2 \sinh(|qH| \rho /4 \pi T^2)} + \frac{4 \pi T^2}{{|qH|}^2 \rho} \Bigg) \nonumber \\
& & \times \, \exp\Big( -\frac{{(M^{\pm}_{\pi})}^2}{4 \pi T^2} \rho \Big) \Bigg[ S\Big( \frac{1}{\rho} \Big) -1 \Bigg] \, .
\end{eqnarray}
The finite-temperature magnetization ${\mathfrak M}(T,M,H)$ -- up to two loops and scaled by $1/T^2$ -- then takes the form
\begin{eqnarray}
\label{magnetizationDensity}
\frac{\mathfrak{M}}{T^2} & = & m_{\pm} h_1(M^{\pm}_{\pi},T,0) + \frac{m_0}{2} \, h_1(M^0_{\pi},T,0) + \frac{t^2}{m^2_H} \,
{\tilde h}_0(M^{\pm}_{\pi},T,H) \nonumber \\
& & + m_{\pm} {\tilde h}_1(M^{\pm}_{\pi},T,H) + {\tilde h}^{[H]}_0(M^{\pm}_{\pi},T,H) \nonumber \\
& & - 4 \pi^2 m^2 \Big\{ 2 m_{\pm} h_2(M^{\pm}_{\pi},T,0) h_1(M^0_{\pi},T,0) + 2 m_0  h_1(M^{\pm}_{\pi},T,0) h_2(M^0_{\pi},T,0) \nonumber \\
& & + 2 m_0 h_2(M^0_{\pi},T,0) {\tilde h}_1(M^{\pm}_{\pi},T,H) + \frac{2 t^2}{m^2_H} \, h_1(M^0_{\pi},T,0) {\tilde h}_1(M^{\pm}_{\pi},T,H)
\nonumber \\
& & + 2 m_{\pm} h_1(M^0_{\pi},T,0) {\tilde h}_2(M^{\pm}_{\pi},T,H) + 2 h_1(M^0_{\pi},T,0) {\tilde h}^{[H]}_1(M^{\pm}_{\pi},T,H) \nonumber \\
& & - m_0 h_2(M^0_{\pi},T,0) h_1(M^0_{\pi},T,0)  \Big\} \, ,
\end{eqnarray}
with coefficients
\begin{eqnarray}
m_{\pm} & = & - \frac{2({\overline l}_6 - {\overline l}_5) m^2_H}{3} \, , \nonumber \\
m_0 & = & - m^2 {\int}_{\!\!\! 0}^{\infty} \mbox{d} \rho \, \rho^{-1} \, \exp\Big( -\frac{m^2}{m^2_H} \rho \Big) \,
\Big( \frac{1}{\sinh(\rho)} - \frac{1}{\rho} \Big) \nonumber \\
& & - \frac{m^4}{m_H^2} \, {\int}_{\!\!\! 0}^{\infty} \mbox{d} \rho \, \exp\Big( -\frac{m^2}{m^2_H} \rho \Big) \,
\Big( \frac{1}{\sinh(\rho)} - \frac{1}{\rho} \Big) \, .
\end{eqnarray}
The dimensionless functions ${\tilde h}^{[H]}_0(M^{\pm}_{\pi},T,H)$ and ${\tilde h}^{[H]}_1(M^{\pm}_{\pi},T,H)$ are
\begin{equation}
{\tilde h}^{[H]}_0(M^{\pm}_{\pi},T,H) = \frac{{\tilde g}^{[H]}_0(M^{\pm}_{\pi},T,H)}{T^2} \, , \qquad
{\tilde h}^{[H]}_1(M^{\pm}_{\pi},T,H) = {\tilde g}^{[H]}_1(M^{\pm}_{\pi},T,H) \, .
\end{equation}

\begin{figure}
\begin{center}
\hbox{
\includegraphics[width=8.0cm]{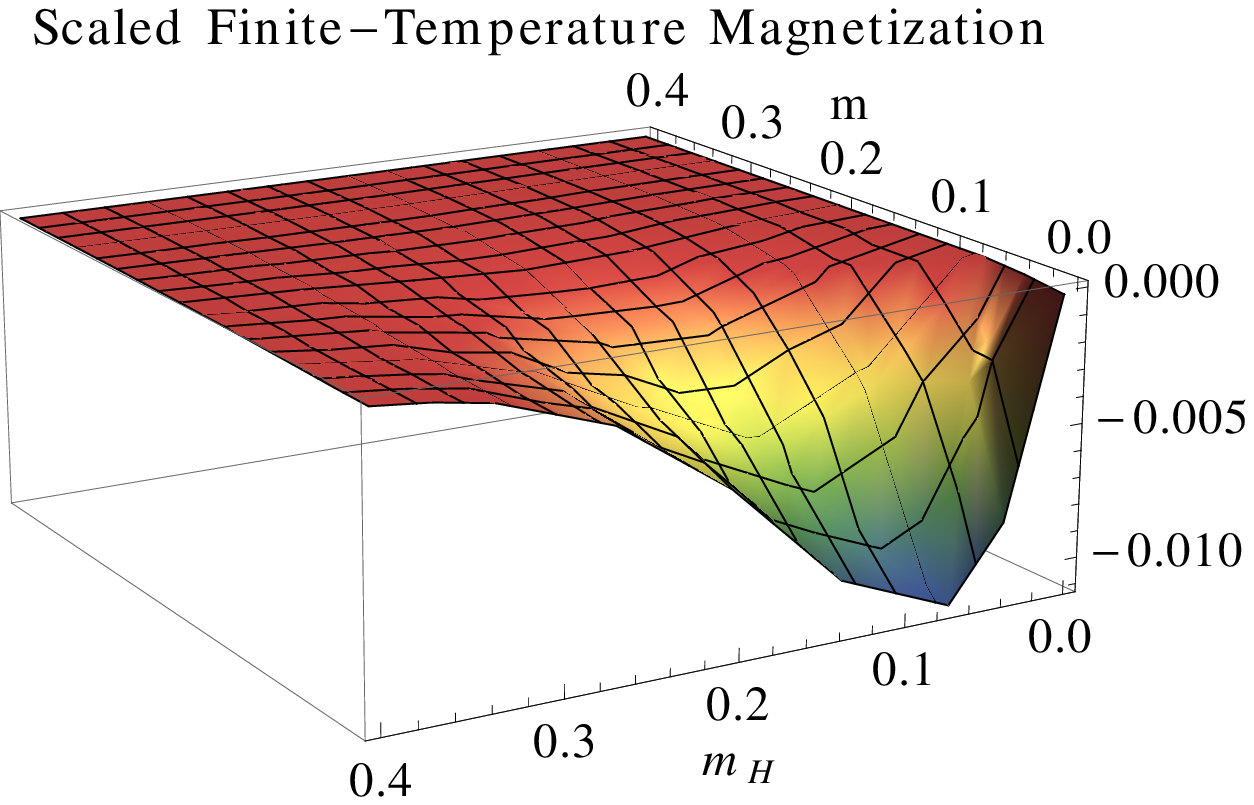}
\includegraphics[width=8.0cm]{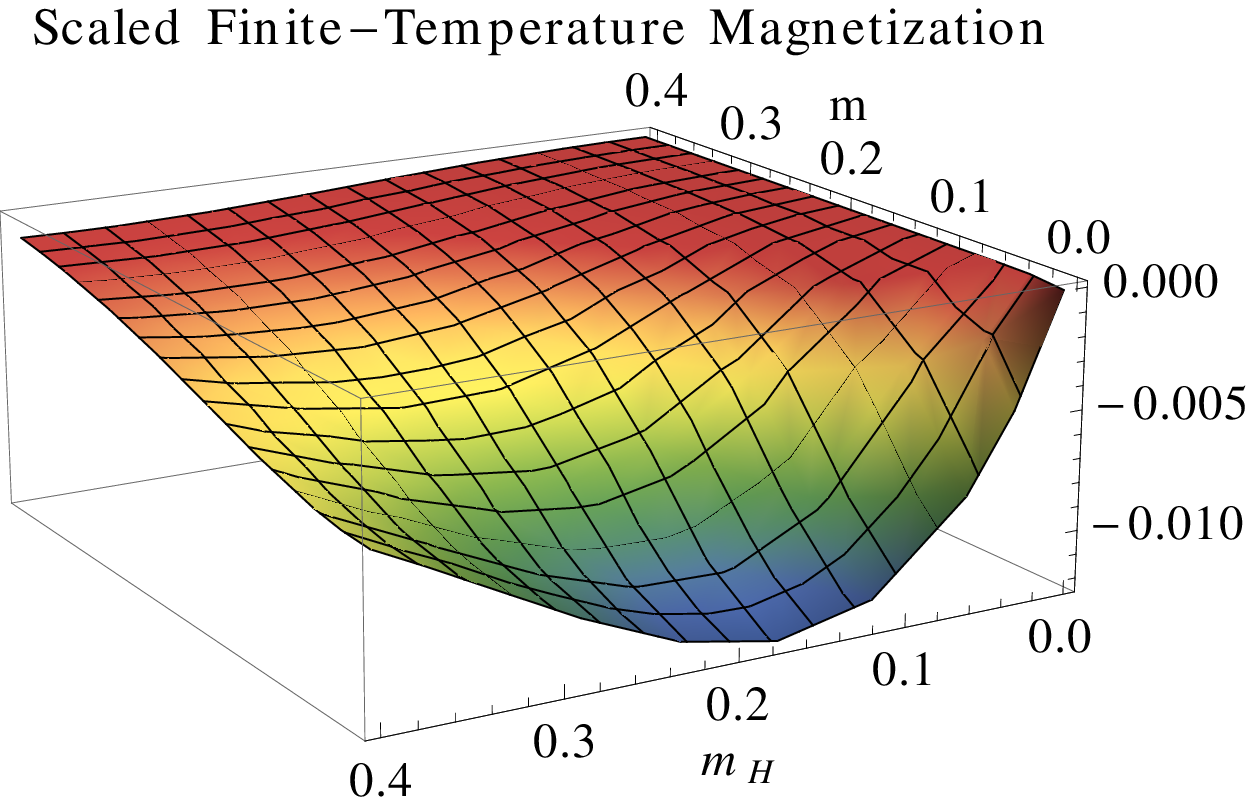}}
\end{center}
\caption{[Color online] Scaled finite-temperature magnetization ${\mathfrak M}/T^2$: Sum of one- and two-loop contributions for
$T= 54 \, \text{MeV}$ (left) and $T= 108 \, \text{MeV}$ (right) in terms of the dimensionless parameters $m,m_H$.}
\label{figure5}
\end{figure}

In Fig.~\ref{figure5} we plot ${\mathfrak M}/T^2$ in terms of magnetic field strength ($m_H$) and pion mass ($m$) at the fixed temperatures
$T= \{ 54, 108 \} \, \text{MeV}$ ($t= \{ 0.05, 0.1 \}$). The finite-temperature magnetization turns out to be negative in the entire
parameter region under consideration. Note that its dependence on magnetic field strength is nontrivial: remarkably, the magnitude of the
finite-temperature magnetization initially grows as the magnetic field increases, but then declines in stronger magnetic fields. For fixed
magnetic field strength, the effect becomes more pronounced as one approaches the chiral limit. On the other hand, the limit $H \to 0$ is
trivial: the finite-temperature magnetization simply vanishes as it should. Accordingly, there is no need to subtract the $H$=0
contribution (as for the entropy density): the quantity ${\mathfrak M}(T,M,H)$ already measures the impact of the magnetic field.

\begin{figure}
\begin{center}
\includegraphics[width=9.5cm]{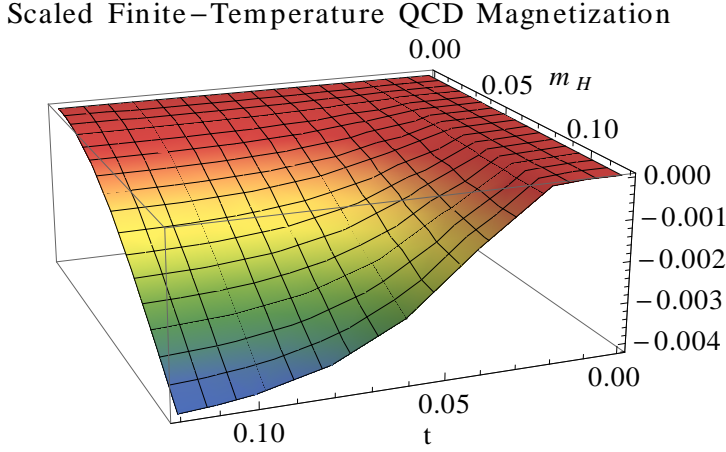}
\end{center}
\caption{[Color online] Scaled finite-temperature QCD magnetization ${\mathfrak M}/T^2$ in terms of the dimensionless parameters $m_H$
(magnetic field strength) and $t$ (temperature).}
\label{figure6}
\end{figure}

In Fig.~\ref{figure6} we explore the dependence of ${\mathfrak M}/T^2$ on magnetic field strength and temperature in the real world with
physical pion mass $M_{\pi} = 140 \, \text{MeV}$ ($m = 0.130$). The induced finite-temperature magnetization is negative and its magnitude
grows as both magnetic field and temperature increase. The dependence of ${\mathfrak M}/T^2$ on the magnetic field -- at fixed temperature
-- implies that the QCD vacuum behaves as a {\it diamagnetic} medium at low temperatures and in weak magnetic fields.

This conclusion fully agrees with studies based on lattice QCD \citep{BEMNS14,BBEKS14,LT14,BEP20}, on the (2+1) flavor Polyakov-loop
quark-meson model \citep{LFL19}, and on the three-flavor quark-meson model with $U_A(1)$ anomaly \citep{KK15}. In these references the
finite-temperature magnetic susceptibility $\chi_B(T)$,
\begin{equation}
\chi_B(T) = \lim_{H\to0} \, -\frac{{\mbox{d}}^2}{\mbox{d} {|qH|}^2} \, \Big\{ z - z_0 \Big\} \, ,
\end{equation}
was considered. At lower temperatures where only pions are the relevant degrees of freedom, the authors indeed find that the magnetic
susceptibility is negative. In particular, its value at $T = 90 \, \text{MeV}$, reported in Ref.~\citep{BBEKS14},
\begin{equation}
\chi_B(T \! = \! 90 \, \text{MeV}) = -0.002(2) \, ,
\end{equation}
is perfectly consistent with our two-loop CHPT analysis where we extract\footnote{We have used $F = 85.6 \, \text{MeV}$ from
Ref.\citep{Aok20} and -- in accordance with Ref.~\citep{End13} -- $M_{\pi} = 135 \, \text{MeV}$.}
\begin{equation}
\chi_B(T \! = \! 90 \, \text{MeV}) = -0.00245 \, .
\end{equation}

\section{Finite-Temperature Quark Condensate}
\label{quarkCondensate}

It is illuminating to include the quark condensate ${\langle {\bar q} q \rangle}$ into discussion and compare its properties at low
temperatures and weak magnetic fields with the behavior of the entropy density. The two-loop representation for the quark condensate,
decomposed into $T$=0 and finite-temperature contributions, 
\begin{equation}
\langle {\bar q} q \rangle = \langle 0 | {\bar q} q | 0 \rangle + {\langle {\bar q} q \rangle}^T \, ,
\end{equation}
has been derived in Ref.~\citep{Hof20b}. Here we just quote the result for the finite-temperature portion ${\langle {\bar q} q \rangle}^T$
that is relevant to make our point,\footnote{The quantity ${\langle 0 | {\bar q} q | 0 \rangle}_0$ is the quark condensate in the chiral
limit at zero temperature and zero magnetic field.}
\begin{equation}
\frac{{\langle {\bar q} q \rangle}^T}{{\langle 0 | {\bar q} q | 0 \rangle}_0} \, {\Big( 1 - \frac{ M^2_{\pi}}{32 \pi^2 F^2} \,
(2 {\overline l}_3 -1)\Big)}^{-1}
= - \Big\{ \frac{q_1}{F^2} T^2 + \frac{q_2}{F^4} T^4  + {\cal O}(T^6) \Big\} \, ,
\end{equation}
with coefficients
\begin{eqnarray}
q_1 & = &  h_1(M^{\pm}_{\pi},T,0) + \mbox{$ \frac{1}{2}$} a_0 h_1(M^0_{\pi},T,0) + {\tilde h}_1(M^{\pm}_{\pi},T,H) \, , \nonumber \\
q_2 & = & + \mbox{$ \frac{1}{2}$} h_1(M^{\pm}_{\pi},T,0) h_1(M^0_{\pi},T,0)
+ \mbox{$ \frac{1}{2}$} h_1(M^0_{\pi},T,0) {\tilde h}_1(M^{\pm}_{\pi},T,H) \nonumber \\
& & - \mbox{$ \frac{1}{8}$} h_1(M^0_{\pi},T,0) h_1(M^0_{\pi},T,0)
- \mbox{$ \frac{1}{2}$} \frac{m^2}{t^2} h_1(M^0_{\pi},T,0) h_2(M^{\pm}_{\pi},T,0) \nonumber \\
& & - \mbox{$ \frac{1}{2}$} a_0 \frac{m^2}{t^2} h_1(M^{\pm}_{\pi},T,0) h_2(M^0_{\pi},T,0)
- \mbox{$ \frac{1}{2}$} a_0 \frac{m^2}{t^2} {\tilde h}_1(M^{\pm}_{\pi},T,H) h_2(M^0_{\pi},T,0) \nonumber \\
& & + \mbox{$ \frac{1}{4}$}a_0 \frac{m^2}{t^2} h_1(M^0_{\pi},T,0) h_2(M^0_{\pi},T,0)
- \mbox{$ \frac{1}{2}$} \frac{m^2}{t^2} h_1(M^0_{\pi},T,0) {\tilde h}_2(M^{\pm}_{\pi},T,H) \, .
\end{eqnarray}
The NLO mass correction $a_0$ reads
\begin{equation}
a_0 = \frac{\mbox{d} {(M^0_{\pi})}^2 }{\mbox{d} M^2_{\pi}} =  1 + \frac{K_1}{F^2}
+ \frac{M^2_{\pi}}{F^2} \, \frac{\mbox{d} K_1}{\mbox{d} M^2_{\pi}} \, ,
\end{equation}
where the integral $\mbox{d} K_1/\mbox{d} M^2_{\pi}$ is
\begin{equation}
\frac{\mbox{d} K_1}{\mbox{d} M^2_{\pi}} = - \frac{1}{16 \pi^2} \, {\int}_{\!\!\! 0}^{\infty} \mbox{d} \rho \, \exp\Big( -\frac{M^2_{\pi}}{|qH|}
\rho \Big) \, \Big( \frac{1}{\sinh(\rho)} - \frac{1}{\rho} \Big) \, .
\end{equation}

\begin{figure}
\begin{center}
\hbox{
\includegraphics[width=8.0cm]{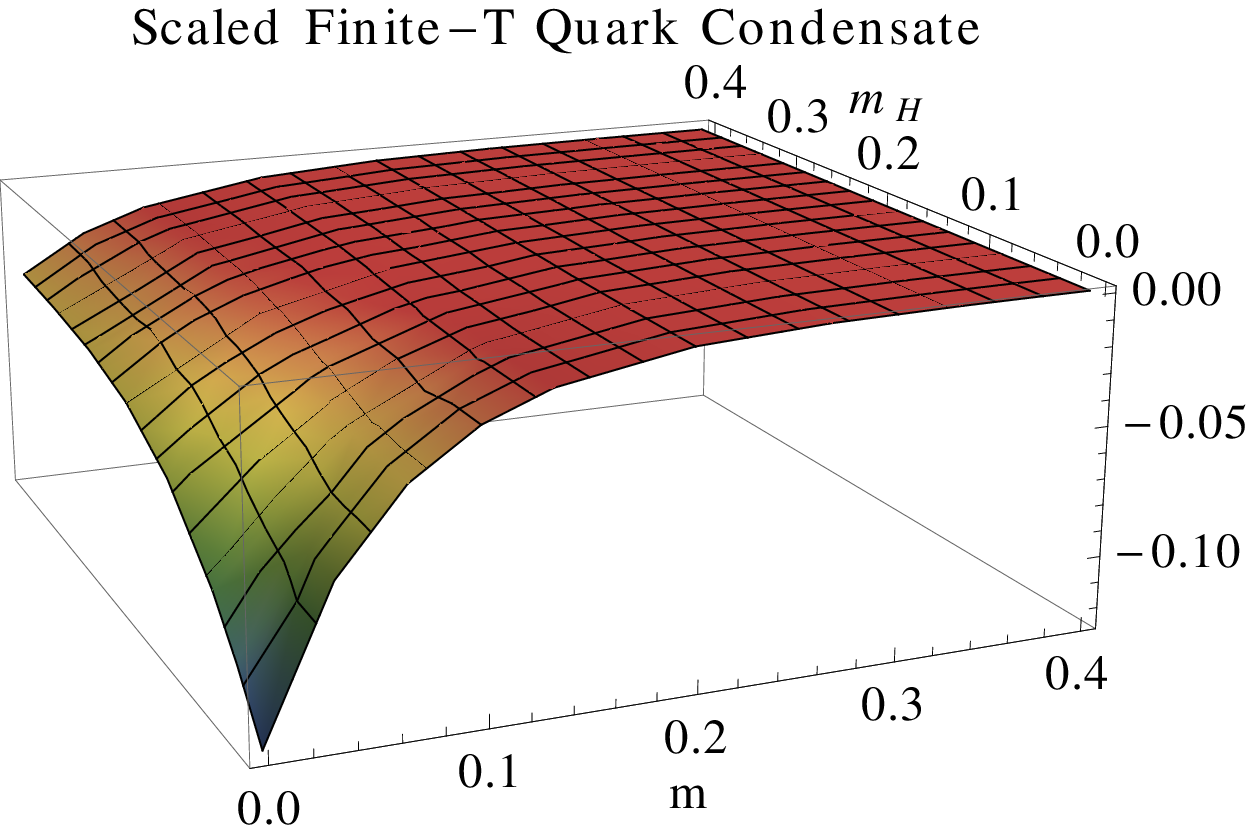}
\includegraphics[width=8.0cm]{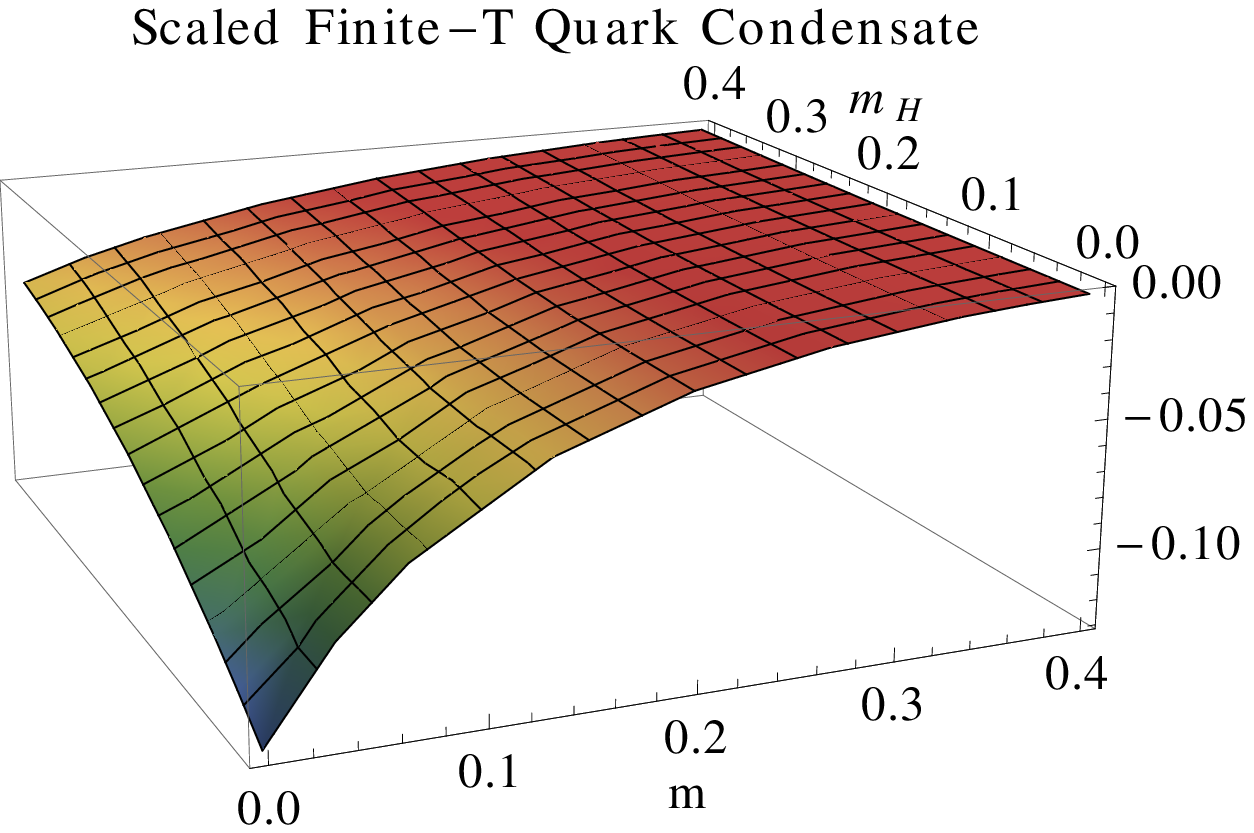}}
\end{center}
\caption{[Color online] Scaled finite-temperature quark condensate: Sum of one- and two-loop contributions for $T= 54 \, \text{MeV}$
(left) and $T= 108 \, \text{MeV}$ (right) in terms of the dimensionless parameters $m$ (pion mass) and $m_H$ (magnetic field strength).}
\label{figure7}
\end{figure}

In Fig.~\ref{figure7} we plot the scaled finite-temperature quark condensate, i.e., the dimensionless quantity
\begin{equation}
- \Big( q_1 + q_2 \, \frac{T^2}{F^2} \Big) \, ,
\end{equation}
in terms of magnetic field strength ($m_H$) and pion mass ($m$) for the two temperatures $T= \{ 54, 108 \} \, \text{MeV}$. The condensate
grows as magnetic field strength or pion masses increase, but the dependence on pion mass is more pronounced.

To better appreciate the impact of the magnetic field -- as for the entropy density -- we subtract the $H$=0 portion and define the
dimensionless quantity
\begin{equation}
\label{condensateH0subtracted}
{\langle {\bar q} q \rangle}^T_H =
\frac{c_0}{T^2} \, \Big\{ {\langle {\bar q} q \rangle}^T - {{\langle {\bar q} q \rangle}^T|}_{H=0} \Big\} \, ,
\end{equation}
where
\begin{equation}
c_0 = \frac{F^2}{{\langle 0 | {\bar q} q | 0 \rangle}_0} \, {\Big( 1 - \frac{ M^2_{\pi}}{32 \pi^2 F^2} (2 {\overline l}_3 -1)\Big)}^{-1} \, .
\end{equation}
Notice that ${\langle {\bar q} q \rangle}^T_H$ captures the effect of the magnetic field in the finite-temperature quark condensate.

\begin{figure}
\begin{center}
\hbox{
\includegraphics[width=8.0cm]{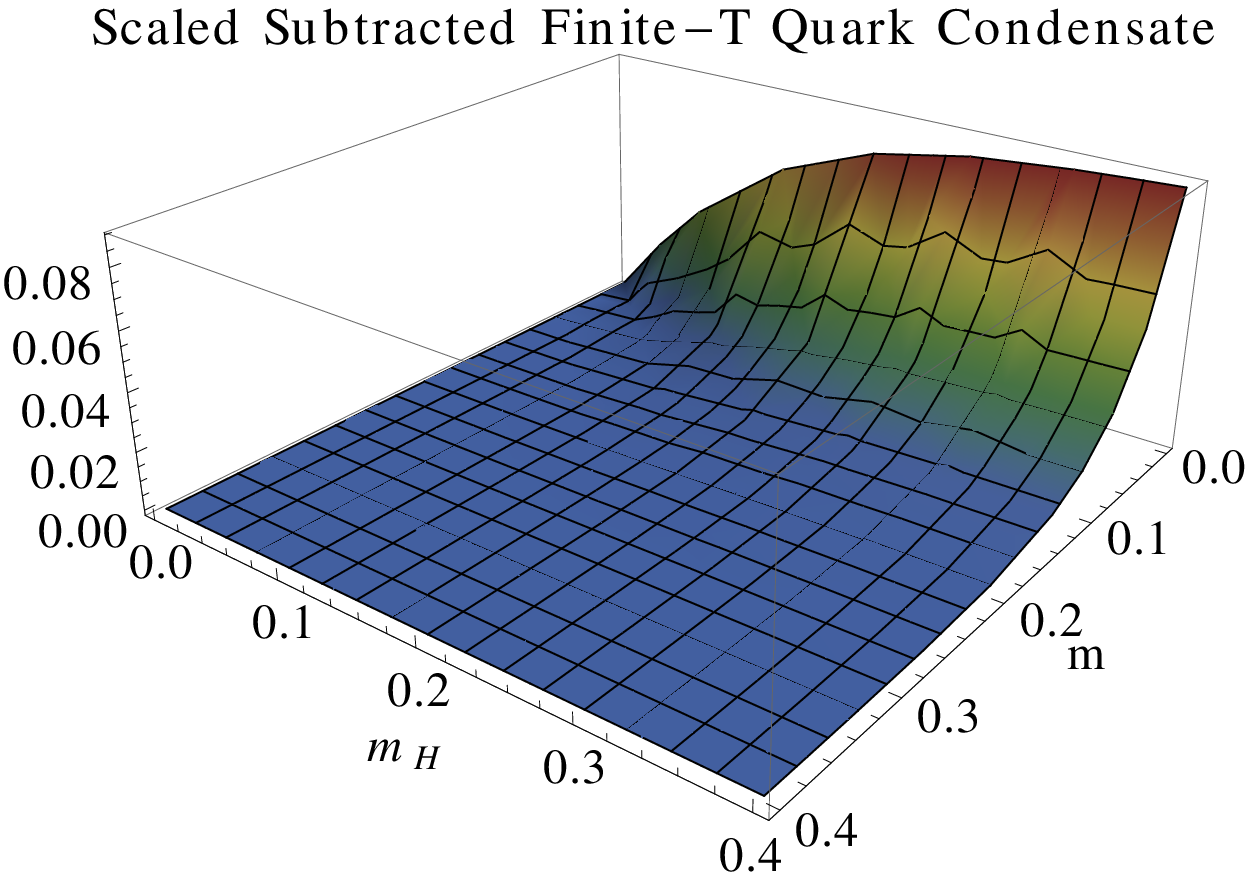}
\includegraphics[width=8.0cm]{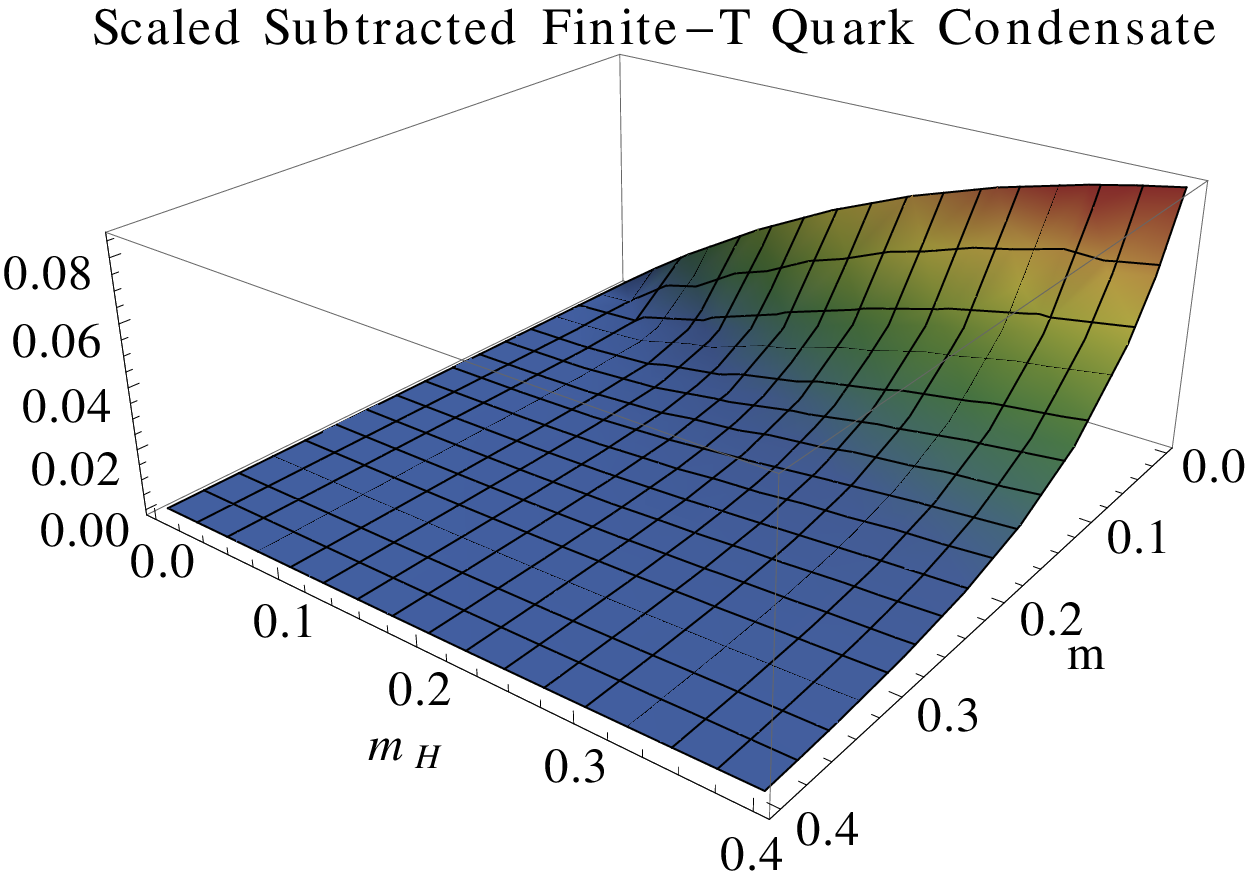}}
\end{center}
\caption{[Color online] Scaled subtracted finite-temperature quark condensate ${\langle {\bar q} q \rangle}^T_H$: Sum of one- and
two-loop contributions for $T= 54 \, \text{MeV}$ (left) and $T= 108 \, \text{MeV}$ (right) in terms of the dimensionless parameters $m$
(pion mass) and $m_H$ (magnetic field strength).}
\label{figure8}
\end{figure}

Inspecting the corresponding Fig.~\ref{figure8}, it now becomes obvious that ${\langle {\bar q} q \rangle}^T_H$ -- i.e., the subtracted
finite-temperature order parameter -- is correlated with the subtracted entropy density (cf. Fig.~\ref{figure2}): the enhancement of the
order parameter caused by the magnetic field is reflected in the decrease of the entropy density. Note that in either subtracted quantity
the influence of the magnetic field is largest in the chiral limit.

\begin{figure}
\begin{center}
\hbox{
\includegraphics[width=8.0cm]{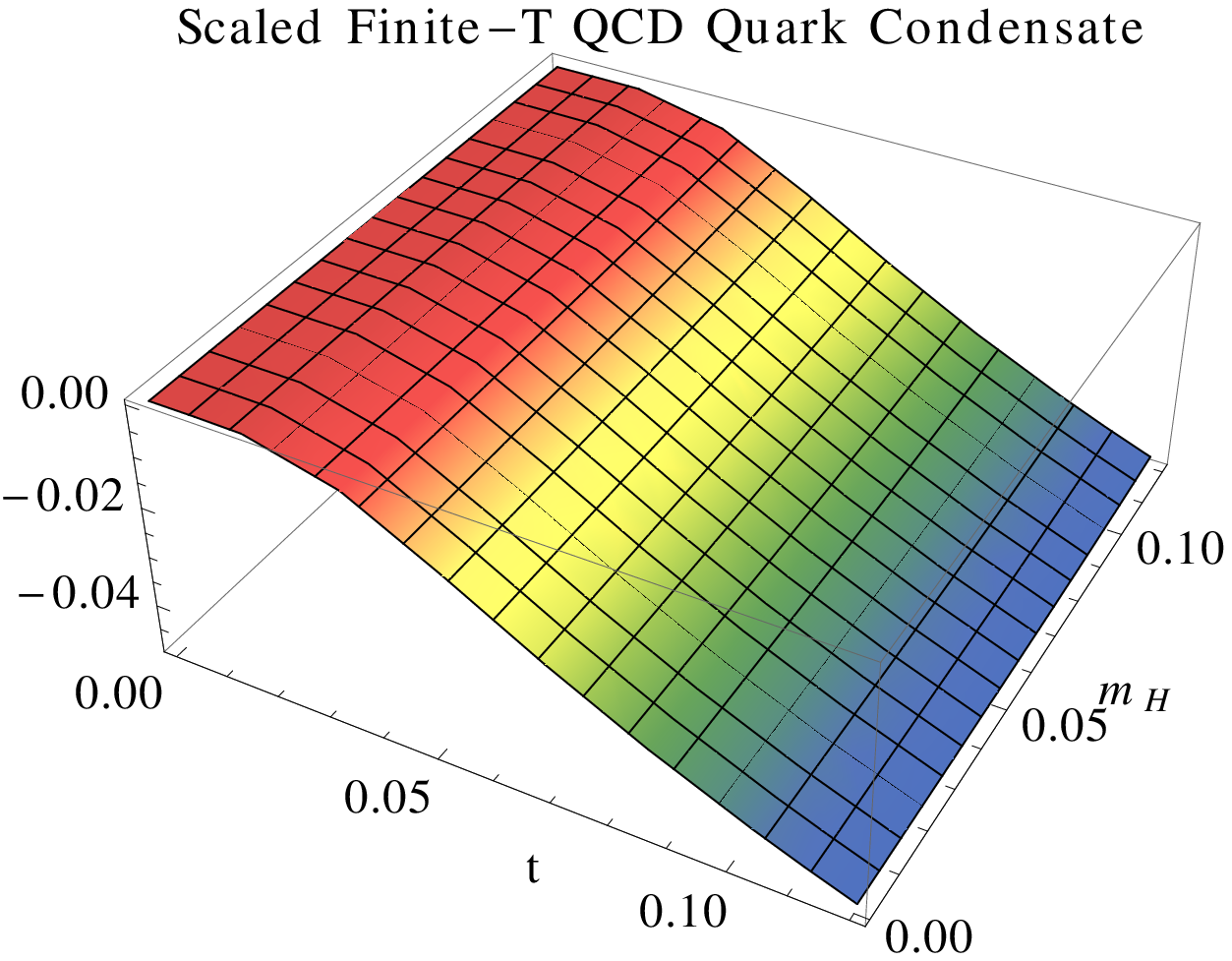}
\includegraphics[width=8.0cm]{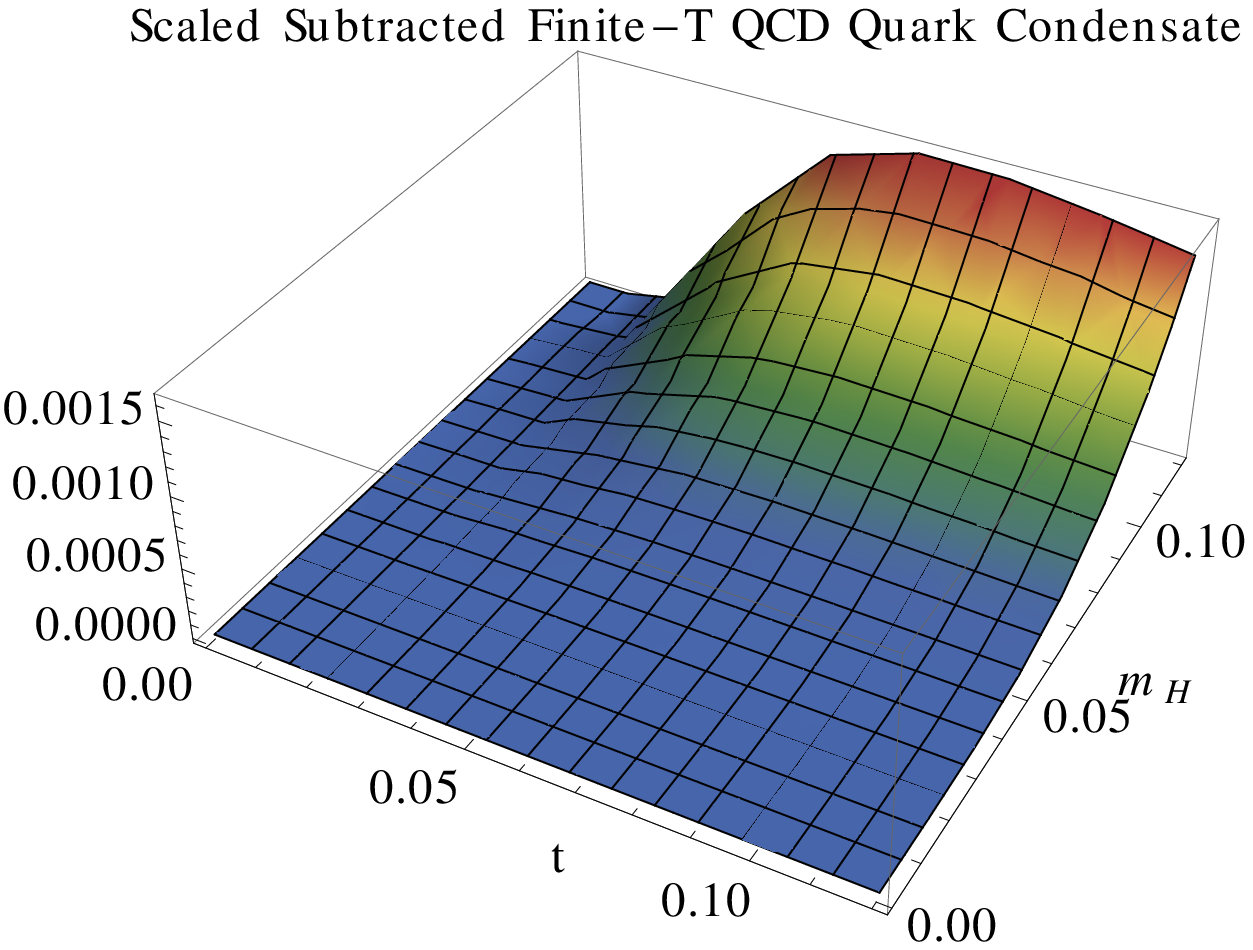}}
\end{center}
\caption{[Color online] Scaled finite-temperature QCD quark condensate in terms of the dimensionless parameters $m_H$ and $t$. LHS: Full
contribution $-(q_1 + q_2 T^2/F^2)$. RHS: Subtracted contribution ${\langle {\bar q} q \rangle}^T_H$.}
\label{figure9}
\end{figure}

Let us finally discuss the connection between quark condensate and entropy density at the physical point $M_{\pi} = 140 \, \text{MeV}$. On
the LHS of Fig.~\ref{figure9} we depict the scaled finite-temperature QCD quark condensate,
\begin{equation}
- \Big( q_1 + q_2 \, \frac{T^2}{F^2} \Big) \, ,
\end{equation}
in function of magnetic field strength and temperature. The influence of the magnetic field is barely visible in comparison to the
predominant dependence on temperature. This motivates us to consider the subtracted QCD quark condensate,
\begin{equation}
{\langle {\bar q} q \rangle}^T_H \, ,
\end{equation}
that we show on the RHS of Fig.~\ref{figure9}. Now the correlation between order parameter and entropy density  (cf. RHS of
Fig.~\ref{figure3}) becomes obvious: the magnetic field enhances the finite-temperature quark condensate which is reflected in the decrease
of the entropy density. It should be noted that the correlation is not one-to-one, but just is of qualitative nature. For example, as
Fig.~\ref{figure9} illustrates, the effect of the magnetic field is most pronounced around $t \approx 0.085$, or
$T \approx 90 \, \text{MeV}$, while for the entropy density the most distinct drop occurs around $T \approx 40 \, \text{MeV}$. 

\section{Conclusions}
\label{conclusions}

Within the framework of two-flavor chiral perturbation theory in a magnetic background, we have derived the two-loop representations for
the entropy density and the magnetization. In various figures we have explored how these quantities depend on temperature, magnetic field
strength and pion mass.

We observe that the entropy density -- at fixed temperature -- drops when the magnetic field becomes stronger and that its decrease is most
pronounced in the chiral limit. At the physical point $M_{\pi} = 140 \, \text{MeV}$, the decrease of the entropy density is most distinct
around $T \approx 40 \, \text{MeV}$. The lowering of the entropy density in an external magnetic field has also been reported in
model-based studies. Here, however, we have provided a fully systematic investigation.

The magnetization at finite temperature is negative in the entire parameter domain accessible by CHPT, i.e.,
$T, M_{\pi} , \sqrt{|qH|} \lessapprox 0.2 \, \text{GeV}$. Its dependence on magnetic field strength is nontrivial: the magnitude of the
finite-temperature magnetization initially grows as the magnetic field increases, but then declines in stronger magnetic fields. This
behavior is most distinct in the chiral limit. In the real world ($M_{\pi} = 140 \, \text{MeV}$) the QCD vacuum behaves as a diamagnetic
medium at low temperatures and weak magnetic fields, in accordance with lattice QCD and model-based studies.

Finally we have addressed the connection between the behavior of the entropy density and the finite-temperature quark condensate in a
magnetic field. The connection becomes most obvious when the $H$=0 pieces in either quantity are subtracted, such that the effect of the
magnetic field is revealed: we then observe that the enhancement of the finite-temperature quark condensate in a magnetic field is
reflected in a decrease of the entropy density.

Although the thermomagnetic properties of quantum chromodynamics in a homogeneous external magnetic field have been explored before by
various authors, a comprehensive investigation of the entropy density and the magnetization in the regime of low temperatures and weak
magnetic fields seems to be lacking. Here we have provided a fully systematic analysis relying on two-flavor chiral perturbation theory.

While we have focused on effects emerging at finite temperature, it would be interesting to also discuss zero-temperature effects. This,
however, requires a detailed analysis of the zero-temperature vacuum polarization that involves the "unphysical" next-to-leading order
low-energy constant ${\overline h}_2$ that yet has to be determined. A detailed investigation including the magnetic susceptibility and the
associated diamagnetic and paramagnetic phases -- as well as an extensive comparison with the pertinent literature -- will be given in a
forthcoming study \citep{Hof20c}. It would also be interesting to extend the present case to three-flavor chiral perturbation theory, i.e.,
to include kaons and the $\eta$-particle. An even more ambitious task is to elevate the present two-loop analysis to the three-loop level.
Corresponding work is in progress. 

\section*{Acknowledgments}

The author thanks J.\ Bijnens and H.\ Leutwyler for correspondence.

\end{document}